\documentclass{aa}  
\usepackage{graphicx}
\usepackage{txfonts}
\newcommand{\beq}{\begin{equation}}
\newcommand{\eeq}{\end{equation}}
\newcommand{\D}{\displaystyle}

\begin{document}
\title{A compact dusty disk around the Herbig Ae star HR~5999 
resolved with VLTI / MIDI\thanks{Based on observations with the ESO
Very Large Telescope Interferometer at Paranal Observatory under programme IDs 
073C-0248 and 073C-0720.}}

\author{Thomas Preibisch\inst{1}, Stefan Kraus\inst{1},  Thomas Driebe\inst{1}, 
 Roy van Boekel\inst{2}, Gerd Weigelt\inst{1}}

   \offprints{Thomas Preibisch}

\institute{Max-Planck-Institut f\"ur Radioastronomie,
 Auf dem H\"ugel 69, D--53121 Bonn, Germany \and 
 Max-Planck-Institut f\"ur Astronomie, K\"onigstuhl 17, 
  D--69117 Heidelberg, Germany
} 

   \date{Submitted: 06/02/2006 ; accepted: 22/06/2006}

 
  \abstract
   {}
   {We have used mid-infrared  long-baseline interferometry 
to resolve the circumstellar material around the Herbig Ae star HR~5999,
providing the first direct measurement of its angular size, and to
derive constraints on the spatial distribution of the dust.}
   {MIDI at the VLTI was used to obtain a set of ten spectrally dispersed 
($8-13\,\mu$m) interferometric measurements of HR~5999 at different 
projected baseline lengths and position angles.
To derive constraints on the geometrical distribution of the dust, we
compared our interferometric measurements to
2D, frequency-dependent radiation transfer simulations of circumstellar
disks and envelopes.
}
   {The derived visibility values between $\sim$0.5 and $\sim$0.9 show
that the mid-infrared emission from HR~5999 is clearly resolved.
The characteristic size of the emission region depends on the 
projected baseline length and position angle, and it ranges between  
$\sim 5-15$~milliarcseconds (Gauss FWHM), corresponding to remarkably 
small physical sizes of $\sim 1-3$ AU. 
For disk models with radial power-law density
distributions, the relatively weak but very extended emission from 
outer disk regions ($\ga 3$~AU) leads to 
model visibilities that are significantly lower than the
observed visibilities, making these models inconsistent with the MIDI data.
Disk models in which the density is truncated at outer radii of 
$\sim 2 - 3$~AU, on the other hand,
provide good agreement with the data.
}
   {A satisfactory fit to
the observed MIDI visibilities of HR~5999 is found
with a model of a geometrically thin disk that is 
truncated at $2.6$~AU and seen under an inclination angle of       
$58\degr$ (i.e.~closer to an edge-on view than to a face-on view).
Neither models of a
geometrically thin disk seen nearly edge-on, nor models of spherical dust 
shells can achieve agreement between the observed and predicted visibilities.
The reason why the disk is so compact remains unclear;
we speculate that it has been truncated by a close binary companion.}

\authorrunning{Preibisch et al.} \titlerunning{VLTI MIDI observations of HR~5999}
   \maketitle

\section{Introduction}

Herbig AeBe stars (HAEBEs; see Herbig 1960; Herbig 1994; Th\'e et al.~1994)
are pre-main sequence stars of intermediate mass
($\sim 2-10\, M_\odot$)
that show broad emission lines, rapid variability, and
excess infrared and millimeter-wavelength emission.
These properties suggest
the presence of substantial amounts of circumstellar dust and gas.
The distribution of the circumstellar material is still a matter of debate.
Since until recently the spatial scales of the inner circumstellar
environment (a few AU, corresponding to $\la 0.1''$) were not accessible
to optical and infrared imaging observations, conclusions drawn on
the spatial distribution of the circumstellar material of the HAEBE stars
were in most cases based entirely on the modeling of the 
spectral energy distribution (SED).
However, fits to the observed SEDs are highly
ambiguous (e.g.~Men'shchikov \& Henning 1994).
For example, the SEDs of some HAEBEs could be equally well explained with
such very different models as 
geometrically thin accretion disks and 
spherical envelopes
(Hillenbrand et al.~1992; Miroshnichenko et al.~1997).
More complex models, like envelopes with cavities
produced by bipolar molecular outflows (Hartmann et al.~1993),
geometrically thin and
optically thick disks surrounded by a spherical envelope (Natta \&
Kr\"ugel~1995, Natta et al.~2001; Miroshnichenko et al.~1999), 
flared outer disks,
puffed-up inner disk rims (Dullemond et al.~2001), and 
disk plus inner halo models (Vinkovic et al.~2006)
have also been used to successfully fit the observed SEDs of HAEBEs.
The strongest evidence for circumstellar disks around HAEBEs comes
from millimeter interferometry (Mannings \& Sargent 1997, 2000): 
 flattened structures
around several sources have been resolved on 
$\sim 100$ AU scales at millimeter wavelengths and
detailed kinematic modeling of one source, MWC 480, is consistent with rotation
in a Keplerian disk.

Speckle interferometry by Leinert et al.~(2001) in the
NIR reveals numerous examples of resolved halos on scales of $\sim 0.1'' - 1''$
(corresponding to several hundred AU).
Bispectrum speckle interferometry 
revealed complex asymmetric
structures on scales down to 70~mas in several intermediate- to high-mass
young stellar objects (e.g. Schertl et al.~2000; Weigelt et al.~2002ab, 2006; 
Preibisch et al.~2001, 2002, 2003; Hofmann et al.~2004).
Interferometric aperture masking Keck observations of
the young system LkH$\alpha$ 101~(Tuthill et al.~2001; 2002) revealed 
some details of the inner disk, which is seen nearly face-on.
  To make
further progress on this topic, observations with even 
higher spatial resolution, at the level $\la 10$~milli-arcseconds (mas),
are needed to explore the
inner circumstellar regions of these objects in detail.
Such observations can only be performed  with long-baseline
interferometry.

The current generation of infrared long-baseline interferometers
is sensitive to the distribution
of dust around the nearest young stars on scales on the order of 1 AU, and
it provides a powerful probe of the models of disks and envelopes of such stars.
The first systematic study of HAEBEs using the technique of long-baseline
interferometry in the near-infrared was conducted by Millan-Gabet et al.~(2001),
who could resolve the near-infrared emission
of 11 of the 15 HAEBEs observed at the IOTA interferometer.
The derived sizes of the observed structures are 0.5 to 6 AU, but
no unique geometrical model could be determined for any of the sources
due to the very sparse coverage of the  $(u,v)$-plane. A number of further studies
have been obtained at different interferometers 
(e.g.,~Akeson et al. 2000; 
Wilkin \& Akeson 2003; Eisner et al.~2004; Monnier et al.~2005).
Most of the near-infrared  interferometric data
 are consistent with a simple disk model
possessing a central dust-free cavity, ringed by hot dust, but other
geometries can often not be excluded, due to the poor sampling of these
observations of the  $(u,v)$-plane.

The first long-baseline interferometric observations of
HAEBEs at mid-infrared wavelengths were recently presented by
Leinert et al.~(2004).
Their data, obtained with MIDI at the VLTI,
trace the $8 - 13\,\mu$m emission from hot and warm
dust in the inner disk regions and can be used to
constrain the geometrical structure of the circumstellar material
on angular scales of $\sim 10$~mas.
They resolved the circumstellar material around all of the seven observed, 
nearby HAEBEs
and derived characteristic dimensions of the emitting regions at $10\,\mu$m
ranging from 1~AU to 10~AU, but, again, they
could not determine the geometry of the
dust distributions due to the limited  $(u,v)$-coverage of the observations.

\medskip

HR~5999 (= HD~144668 = V856~Sco) is 
one of the best studied HAEBEs.
It lies in the central part of the Lupus 3 dark cloud, which is part
of the extended Lupus star-formation complex. HR~5999 has a spectral type 
of A7~III-IV and  forms a $\sim 45''$ wide proper-motion binary
system with the A1.5 star HR~6000.
The Hipparcos parallax for HR~5999 corresponds to a distance 
of $208^{+50}_{-30}$~pc.
The spectral energy distribution of HR~5999 shows a clear infrared excess 
at wavelengths above $1\,\mu$m. IRAS detected HR~5999
at 12, 25, and 60$\,\mu$m, but provided only an upper limit
at 100$\,\mu$m. Henning et al.~(1994) found weak millimeter emission
from HR~5999, but the detection was only at the $1\,\sigma$ significance
level.
The far-infrared and mm-emission of HR~5999 suggests the total
mass of the circumstellar material to be on the order 
$0.006\,M_\odot$ (Siebenmorgen et al.~2000).
HR~5999 belongs to the ``group I objects'' as defined by Hillenbrand (1992);
the infrared excess of these objects is thought to be from emission arising 
in a flat, optically thick accretion disk.
The fit of the observed SED with a simple analytical model of an
optically thick circumstellar disk yielded an outer radius of $\ga 23$~AU
and a possible inner hole of $\sim 0.1$~AU.
Siebenmorgen et al.~(2000) analyzed mid-infrared ISO spectra
of HR~5999 and found a broad silicate emission
feature with a peak around $9.6\,\mu$m. With a ratio of peak- to 
continuum-flux of about 1.4, the intensity of this feature
is relatively small. The broad and rather flat shape of the silicate 
feature indicates
the presence of relatively large ($\ga 1\,\mu$m) silicate grains 
(cf. van Boekel et al.~2003).

The stellar parameters of HR~5999 suggest that the object is
$\sim 0.5$~Myr old and has a mass of $\sim 3-4\,M_\odot$ 
(van Boekel et al.~2005).
HR~5999 will thus arrive at the main sequence as a mid B-type star.
HR~5999 shows only moderate extinction  ($A_V = 0.49$~mag) and
is a fast rotator
($v \sin i = 204$~km/sec; Royer et al.~2002). 
The quasi-periodic and irregular photometric and
spectroscopic variability of HR~5999  in the optical 
(see Perez et al.~1992) and in the UV range (see Perez et al.~1993) 
has been interpreted as  
due to instabilities in an optically thick circumstellar accretion disk.
The detailed analysis of the UV emission
lines in the IUE spectra of HR~5999 by Perez et al.~(1993) 
revealed gas
accreting onto the star with velocities as high as 300 km/sec
(see also Blondel et al.~1993). Perez et al.~(1993) also concluded that the
observed double-peaked H$\alpha$ and Mg~II
emission profiles strongly suggest 
 that the circumstellar disk around HR~5999 is seen nearly edge-on.
The average mass accretion rate was estimated to be 
$\sim 7 \times 10^{-7}\,M_\odot/{\rm yr}$, but the observed 
variability of the UV emission clearly suggests that the accretion
process happens in a non-steady fashion.

To summarize, the existing data clearly show that HR~5999 is surrounded
by circumstellar material, probably in the form of an
accretion disk. The high $v \sin i$ and the double-peaked 
structure of the UV emission lines
suggest that the system is seen under a relatively high inclination angle
(i.e.~close to edge on). This makes it an interesting target for a detailed
interferometric study of the geometry of the circumstellar dust.
Grady et al.~(2005) presented HST STIS white-light coronographic imaging
results for HR~5999, found no indication of nebulosity at $r > 70$~AU, and
concluded that the disk around HR~5999 must be geometrically small.

HR~5999 has a close visual companion, Rossiter~3930 (see~Stecklum et al.~1995,
and references therein), with $\Delta V \sim 4.6$ and $\Delta K \sim 3.6$,
 at a projected
separation of $1.4''$,  corresponding to a projected distance
of about 300~AU. Since Rossiter 3930~shows strong H$\alpha$ emission, it 
is very likely a T~Tauri star.
If Rossiter~3930  is actually physically related to HR~5999 and not 
a chance projection, it may be responsible for the rather low
\mbox{(sub-)mm} emission and corresponding low mass of the circumstellar material
of HR~5999 due to the mechanism of dynamic disk clearing by 
gravitational binary interaction (e.g.~Arymowicz \& Lubow 1994).

\section{Observations and data analysis}

\begin{table}
\centering
\caption{Log of MIDI observations of HR~5999. 
}
\label{table_obs}       
%
%
\begin{tabular}{lrrc}
\hline\noalign{\smallskip}
UT date \& time & B & PA & Obs. \\
 & \hspace{0.6cm}[m] & \hspace{0.3cm} [deg] & name\\
\noalign{\smallskip}\hline\noalign{\smallskip}
2004-04-10 \, 02:44:34 &  46.34 & 173.37 & 46m/173\degr\\
2004-04-10 \, 05:23:56 &  46.60 &  21.57 & 46m/21\degr\\
2004-04-11 \, 05:16:58 &  46.59 &  20.74 & 46m/20\degr\\
2004-04-11 \, 09:09:04 &  42.42 &  52.12 & 42m/52\degr\\
2004-04-12 \, 09:46:20 &  39.76 &  56.36 & 39m/56\degr\\
2004-06-28 \, 00:15:43 & 102.38 &  15.35 &\,\,\,102m/15\degr\\
2004-06-28 \, 02:06:06 & 100.76 &  30.90 &100m/30\degr\\
2004-06-28 \, 04:26:08 &  90.89 &  46.10 & 90m/46\degr\\
2004-06-28 \, 05:22:28 &  83.22 &  50.57 & 83m/50\degr\\
2004-09-30 \, 00:02:18 &  63.50 & 123.74 & 63m/123\degr\\
\noalign{\smallskip}\hline
\end{tabular}
\end{table}

\begin{table}
\begin{center}
\caption {
List of calibrators used for the calibration of the HR~5999 data 
together with uniform-disk diameters ($d_{\rm{UD}}$) 
and the date, as well as the time stamp ($t_{\rm obs}$) 
of MIDI observations. 
}
\begin{tabular}{r r c l}\hline
Calibrator & $d_{\rm{UD}}$ (mas) & night  & $t_{\rm obs}$ (UTC)\\
HD number  &                     & (2004) &                    \\ \hline
\hline
 81797     & $9.54\pm0.75$       & 2004-04-10 & 00:41:19       \\
           &                     & 2004-04-11 & 03:56:19       \\ \hline
107446     & $4.54\pm0.23$       & 2004-04-10 & 02:13:31       \\
           &                     & 2004-04-11 & 23:55:03       \\
           &                     & 2004-04-12 & 01:24:27       \\ \hline
109379     & $3.30\pm0.17$       & 2004-06-27 & 23:27:48       \\ \hline
129456     & $3.39\pm0.09$       & 2004-04-10 & 05:46:05       \\
           &                     & 2004-04-12 & 08:37:42       \\
           &                     & 2004-04-12 & 10:09:08       \\
           &                     & 2004-06-28 & 00:43:13       \\ \hline
134505     & $2.55\pm0.13$       & 2004-06-28 & 01:43:12       \\ \hline
139127     & $3.30\pm0.22$       & 2004-04-11 & 05:35:48       \\
           &                     & 2004-04-11 & 09:36:42       \\
           &                     & 2004-04-12 & 07:09:25       \\
           &                     & 2004-06-28 & 02:48:50       \\
           &                     & 2004-06-28 & 05:43:29       \\ \hline
152885     & $2.88\pm0.09$       & 2004-04-10 & 07:13:48       \\
           &                     & 2004-04-12 & 05:57:26       \\
           &                     & 2004-04-12 & 09:24:41       \\ \hline
161892     & $3.94\pm0.21$       & 2004-04-12 & 04:42:03       \\
           &                     & 2004-04-12 & 05:02:42       \\ \hline
168723     & $2.88\pm0.13$       & 2004-04-10 & 09:29:56       \\
           &                     & 2004-04-10 & 10:12:13       \\
           &                     & 2004-09-30 & 01:22:01       \\ \hline
169916     & $3.39\pm0.09$       & 2004-04-12 & 07:57:12       \\ \hline
176411     & $2.24\pm0.13$       & 2004-04-10 & 08:08:34       \\
           &                     & 2004-04-10 & 08:45:41       \\ \hline
188512     & $2.07\pm0.10$       & 2004-09-30 & 03:13:45       \\ \hline
\label{table_calib}
\end{tabular}
\end{center}
\end{table}

\begin{figure}
\begin{center}\includegraphics[width=8cm]{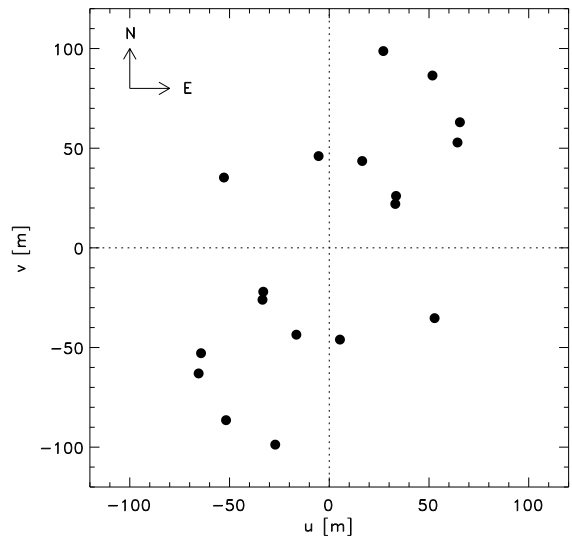}\end{center}
\caption{Points in the $(u,v)$ plane sampled by our MIDI observations
of HR~5999.
} \label{uv}%
\end{figure}

The observations discussed in this paper
were obtained  with the MIDI instrument at the VLTI during several
observing runs between April and September 2004. 
In total, ten MIDI measurements of HR~5999 were performed:
six visibility measurements for observing program
073C-0248 (PI: Preibisch) and four visibility measurements
for program 073C-0720  (PI: van Boekel).
The combined data set is analyzed in this paper.
The observation dates, the projected baseline lengths,
the position angles of the projected baseline on the sky, and the
abbreviated data set names used in the following text
are summarized in Table~\ref{table_obs}.
Our observations cover a range of
projected baseline lengths from 39~m to 102~m
and position angles ranging from 15\hbox{$^\circ$} to 173\hbox{$^\circ$}, 
and, therefore,
provide  a relatively good coverage of the $(u,v)$ plane.

The fringes were scanned by applying rapid internal optical path difference 
(OPD) changes of a few wavelengths around the point of zero OPD. 
A prism with a spectral resolution of $\lambda/\Delta \lambda \simeq 30$ 
was used to obtain spectrally dispersed fringes in the range 
of 8 to 13~$\mu$m.  Immediately after a recording of
interferometric data, photometric calibration data were obtained by 
blocking each of the two beams alternately.  
Chopping with a frequency of 2~Hz was applied 
to obtain the data for the sky background subtraction.
A more detailed description of the observing procedure is given in 
Przygodda et al. (\cite{przygodda03}), \cite{leinert04}, 
Chesneau et al. (\cite{chesneau05a}, \cite{chesneau05b}), 
and Ohnaka et al. (\cite{ohnaka05}).  

Calibration stars with known angular diameter were observed 
for the calibration of the raw visibilities. 
Table~\ref{table_calib} gives an overview of 
these calibration stars, their uniform-disk diameters as given in 
the CalVin list\footnote{available from ESO at
http://www.eso.org/observing/etc/},
and the MIDI 
observations that we used for the calibration 
of the raw visibilities of HR~5999.

We utilized two different MIDI data reduction packages: MIA developed 
at the Max-Planck-Institute for Astronomy and EWS developed 
at the Leiden Observatory.   While the MIA package is based on 
the power spectrum analysis and measures the total power 
of the observed fringes, the EWS package first corrects for OPDs 
(instrumental, as well as atmospheric delays) in each scan, and 
then the interferograms are averaged.
A more detailed description of the MIA and EWS packages 
can be found  in Leinert et al.~(2004)  and 
Jaffe (\cite{jaffe04}), respectively. 
In our analysis of the HR~5999 data, we
found good agreement between the results of both packages 
within the error bars.
In the following discussion we  always refer only
to the results obtained with the MIA software.

The errors of the calibrated visibility were estimated from
 the $1\,\sigma$ scatter of the visibilities calibrated with all individual
   data sets of a night. In addition, to account for systematic error
   contributions (e.g.~due to imperfect beam overlap), we assumed a
   minimum visibility error of 15\%.
One of the ten visibility measurements, the 46m/$21\degr$ data set,
was affected by strong variations in the transfer function around the time 
of the measurement. This particular data set could not be calibrated well;
 we decided to drop it from the
  following analysis.

\section{Observed visibilities}


\begin{figure}\begin{center}
\includegraphics[width=8.4cm]{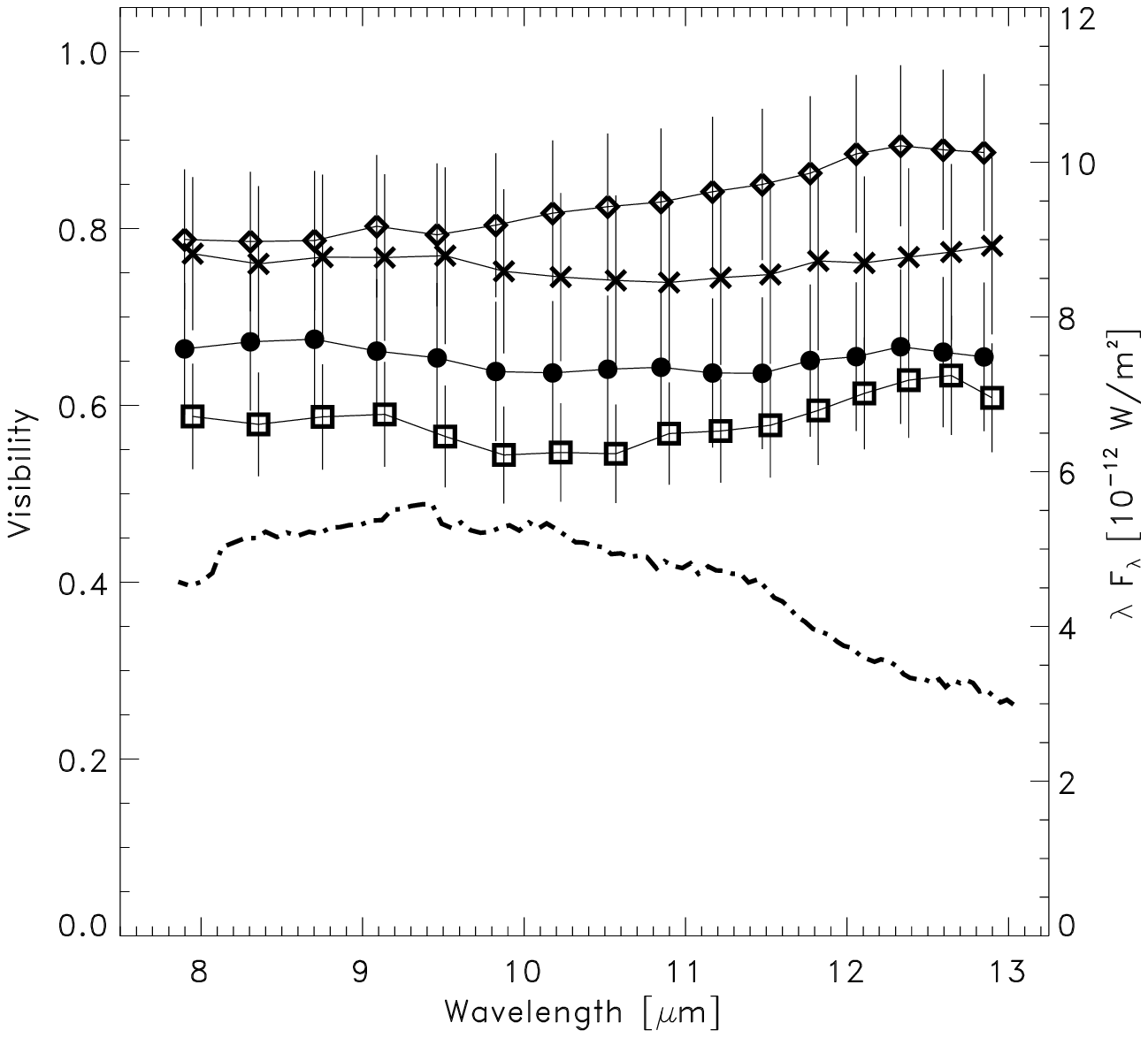}
\includegraphics[width=8.4cm]{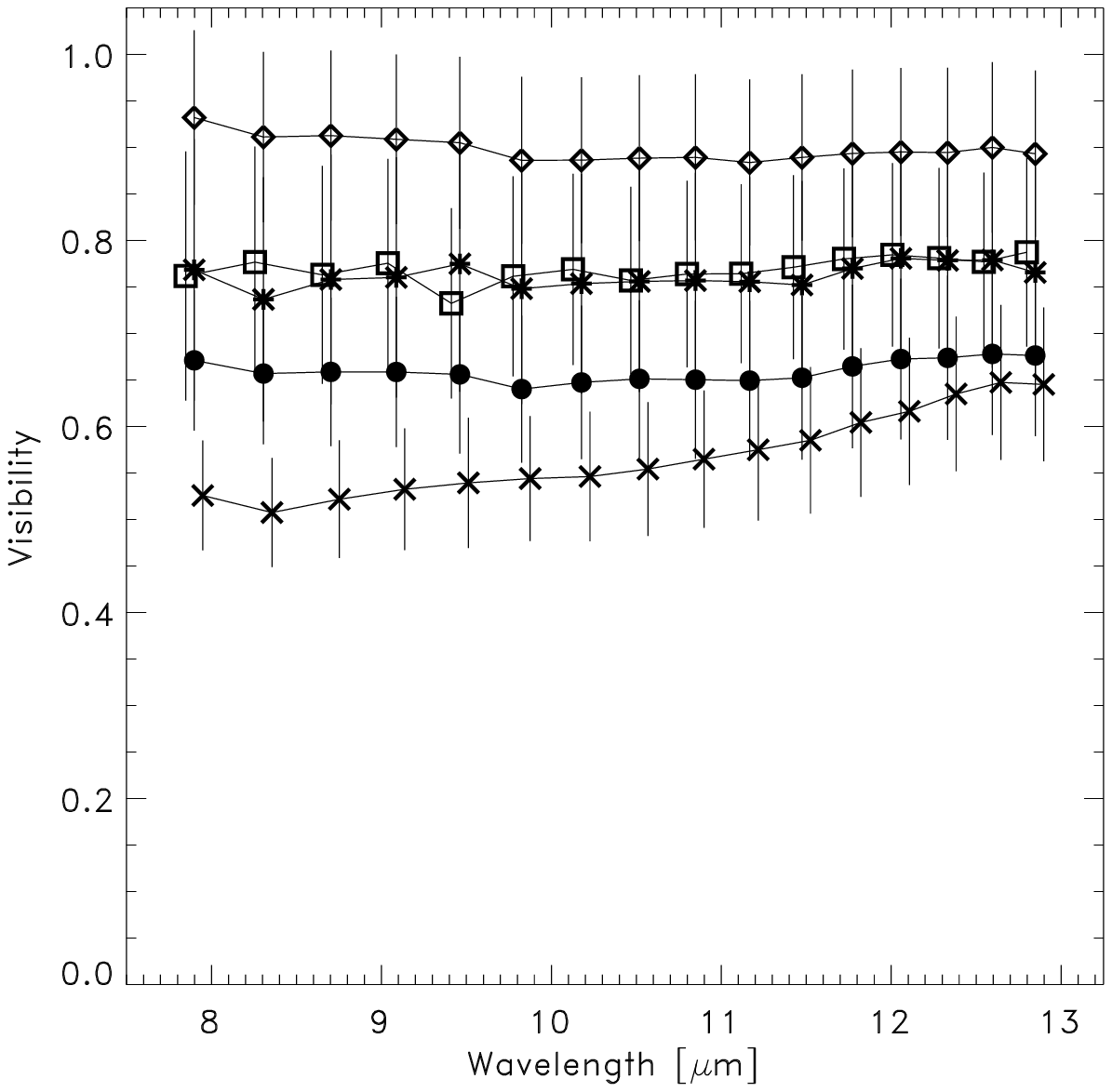}\end{center}
\caption{MIDI visibilities of HR~5999 plotted versus wavelength.
The upper panel shows the data for the observations at
baseline-length\,/\,position angles of
102m/$15\degr$ (dots), 83m/$50\degr$ (crosses), 63m/$123\degr$ (boxes), 
 and 46m/$20\degr$ (diamonds).
The lower panel shows the data for the observations at 100m/$30\degr$ (dots), 90m/$46\degr$ (crosses),
46m/$173\degr$ (boxes), 42m/$52\degr$ (diamonds), and 39m/$56\degr$ (asterisks).
The thick dash-dotted line  and the corresponding scale on the right y-axis
in the upper panel show the flux-calibrated 
spectrum obtained with MIDI during the 
39m/$56\degr$ observation.
} \label{vis}%
\end{figure}

As the first step of the analysis, we consider the dependence of the
observed visibilities on the wavelength, the projected baseline length,
and the baseline position angle. Figure~\ref{vis} shows 
the  wavelength-dependence of the visibilities of HR~5999.
All visibility curves are rather flat,
i.e.~the observed visibilities show no strong dependence 
on the wavelength.
\begin{figure}
\begin{center}\includegraphics[width=8.0cm]{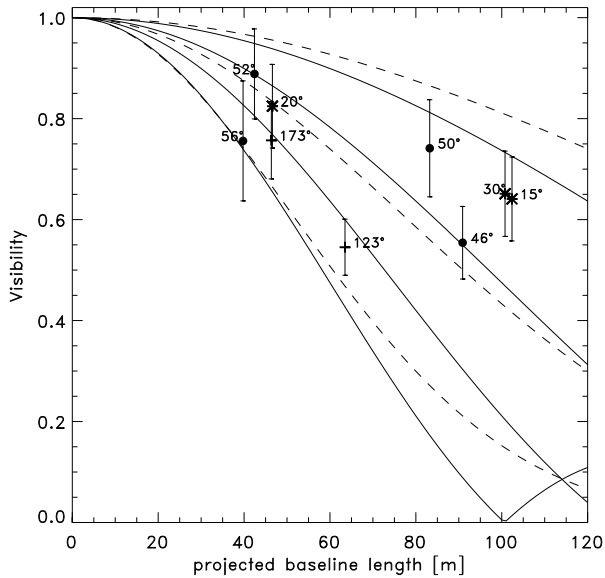}\end{center}
\caption{Visibility at $10.5\,\mu$m as a function of projected baseline length.
The data points are annotated with their respective position angles.
For comparison, we also plotted the theoretical visibility curves
for uniform disk models with diameters of 10, 15, 20, and 25~mas
(solid lines) and for Gauss models with FWHM = 5, 10, and 15~mas
(dashed lines).
} \label{vis_rbase}%
\end{figure}

Figure \ref{vis_rbase} shows the observed $10.5\,\mu$m visibilities
 as a function of projected baseline length.
One can see that the data are not consistent with simple uniform-disk
or Gaussian brightness distribution models.

\subsection{Size of the emission region}\label{size}

\begin{figure*}
\includegraphics[width=12.0cm]{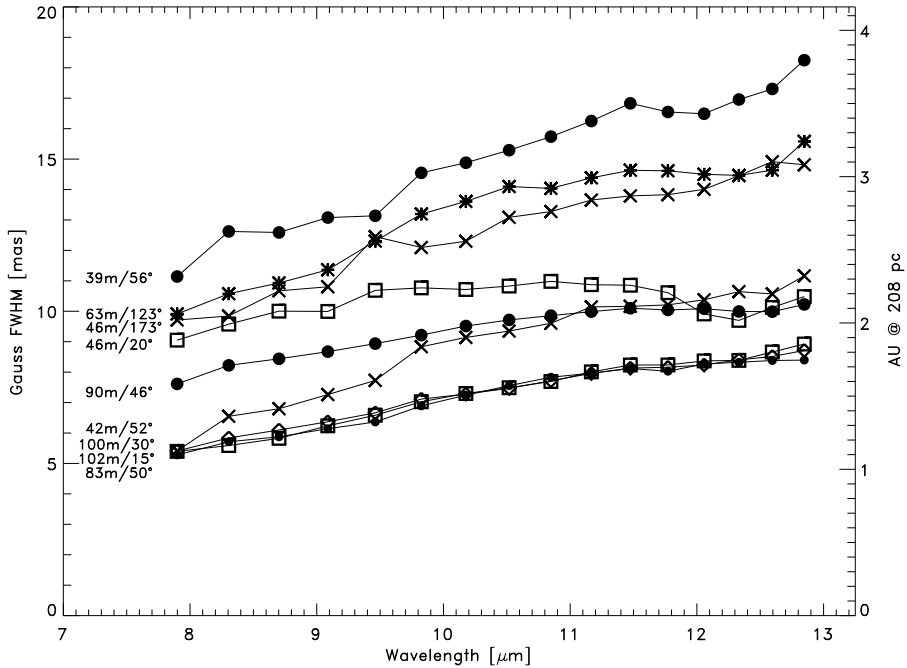}
\caption{Gauss-fit FWHM diameters
as a function of wavelength derived for the MIDI observations of HR~5999.
The individual observations are identified by their values for
projected baseline length and position angle.
} \label{gauss}%
\end{figure*}

In order to derive an initial rough estimate of the size of the 
resolved emission region, we fitted each observed visibility value with
two simple models: first, a uniform-disk model, and second,
a Gaussian brightness distribution.
The resulting diameters range between
10 and 25 milliarcseconds (corresponding to $\sim 2-5$ AU at the distance
of 208~AU) for the
uniform disk model and $5-15$~mas ($\sim 1-3$ AU) for the Gauss model
(FWHM).
The Gauss-fit diameters as a function of wavelength for the different
MIDI observations are shown in Fig.~\ref{gauss}.
For all data sets, the  diameters increase with increasing wavelength
in the MIDI band. This effect can be understood as a consequence of the
fact that the emission at longer wavelengths comes from cooler material,
which is located at larger distances from the central star than the 
warmer material radiating at shorter wavelengths.
Typically, the diameters at $13\,\mu$m are $\sim 50\% - 100\%$ larger
than at $8\,\mu$m.

Another notable effect is the trend that the longest baselines give
the smallest  diameters. Although 
one must not ignore that the different observations
were obtained at different position angles (see next paragraph), 
this suggests a complex geometry for the emission region, 
as discussed in the next section.

\subsection{Shape of the emission region}\label{shape}

\begin{figure}
\begin{center}\includegraphics[width=8.0cm]{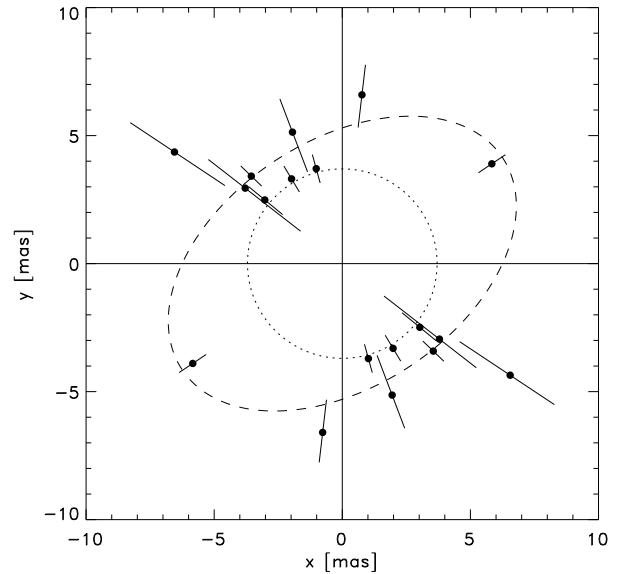}\end{center}
\caption{Gauss-fit FWHM diameters at $10.5\,\mu$m for the different observations
as a function of baseline position angle. The dotted circle has a radius 
of 3.7~mas. The dashed ellipse has a minor axis of 4.8~mas and a major axis of
7.5~mas.
} \label{polar}%
\end{figure}

In Fig.~\ref{polar} we plot the Gauss-fit radii 
(at wavelength 10.5\,$\mu$m)  as
a function of the position angle of the observation.
The plot suggests that the structure of the emission region is not described 
well by a spherically symmetric distribution, i.e.~seems to be
asymmetric. An inclined ellipse with axis ratio 1:1.56 apparently 
yields better agreement
to the data than a circle, but does not provide a particularly good fit either.
The data suggest that the emission has a more complicated
structure, probably with substructure at different size scales.
A more detailed analysis of size and shape of the emission region 
is described below in Section~\ref{rad}. 


\subsection{Comparison to MIDI data of other HAEBEs}\label{leinert}

The visibilities measured for HR~5999 differ from those obtained
by Leinert et al.~(2004) for a sample of seven HAEBEs in two ways.
First, all MIDI measurements of HR~5999 give visibilities $\ge 0.5$,
whereas six of the seven HAEBEs in the Leinert et al.~(2004) sample show
much lower visibility values, $\le 0.4$ in the 
$8\,\mu$m to $13\,\mu$m wavelength range.
Only 51~Oph shows visibility values similar to what we find for
HR~5999.
This implies that
the angular extent of the mid-infrared emission from HR~5999 is 
{\em smaller} than that of most HAEBEs in the Leinert sample.

The second difference is the wavelength dependence of the visibilities.
While the visibility for HR~5999 is essentially flat over the 
$8\,\mu$m to $13\,\mu$m wavelength range, 
most of the HAEBE stars in the
Leinert sample show a rather strong drop in  the visibility between
$8\,\mu$m and $10\,\mu$m, in some cases followed by a slight increase towards
longer wavelengths. The latter wavelength dependence of the visibility
is expected for objects with significant $10\,\mu$m silicate emission features
(a strong emission feature in the spectrum produces
a minimum in the visibility due to optical depth effects, because the 
apparently larger object size in the center of the emission line 
yields lower visibilities).
The flat visibility curves of HR~5999 are probably related to
its very weak silicate emission feature, which does not contribute enough
to produce a significant dependence of the visibility
(apparent object size) on wavelength over this feature.

\section{Constraints on the source shape from 
2D radiative transfer simulations}\label{rad}

The true morphology of the mid-infrared emission
is most likely more complex than  simple geometrical models for the
brightness distribution, like
uniform-disk or Gauss models. 
For example, the existence of a dust-free hole inwards of the dust sublimation 
radius already implies that the brightness distribution cannot have a 
steadily decreasing profile.
In order to see how much information on the spatial distribution
of circumstellar material can be deduced from our
MIDI data, we performed 2D radiation transfer calculations.
We believe that this approach provides more realistic constraints 
than the use of simple analytical models.
Our radiative transfer modeling also includes a self-consistent 
determination of the dust sublimation
radius, i.e.~the inner disk edge, which is a crucial ingredient in any
realistic modeling attempt.
Note that the aim of the modeling in this paper is an exploration 
of different geometries for the dust distribution and not a detailed
disk modeling (which is not warranted considering the rather large
uncertainties of the measured visibilities).

\subsection{Modeling procedure}

For our simulations we used the 2D radiation transfer code described 
in detail by Sonnhalter et al.~(1995).  In this code, 
the distribution of dust temperatures and radiation intensities in
an axially symmetric dusty circumstellar environment around a central
radiation source is calculated within 
the framework of the flux-limited diffusion approximation (Levermore \&
Pomraning 1981). 
The equations are discretized on a quadratic, equally spaced grid.
As the density distribution is assumed to be axially
symmetric with respect to the $z$-axis and mirror symmetric with respect 
to the plane $z=0$,
cylindrical coordinates $(r,z,\phi)$ are used, because the symmetry
reduces the relevant information to the first quadrant of the $r-z$
plane.
To improve resolution and convergence, a system of 5 nested grids with
decreasing grid spacing is used.
The differential
equations for the radiation field are iterated together with the temperature
equations until a self-consistent equilibrium
configuration is reached.
At the end of the iteration procedure, the total radiation flux
(integrated over all frequencies) through the outermost grid cells
was always within 1\% of the stellar luminosity for all 
computed models.
We note that any non-spherical density distribution results in
non-spherically symmetric temperature distributions. For example,
in models of a circumstellar disk, the disk surface is generally
found to be warmer than the disk midplane at the same radial 
distance from the star.

After determination of the dust temperatures for a multi-component dust model,
a ray-tracing procedure is used to calculate intensity maps
for the appearance of the
central object and its circumstellar environment at varying inclinations
for selected frequencies. The simulated images are then used to
compute visibility maps in the $(u,v)$ plane, which finally can be
compared to the MIDI measurements.
We used 64 wavelength points between $0.1\,\mu$m and
5~mm for our modeling of HR~5999.
 The size of the numerical grid was set to $50 \times 50$~AU,
which resulted in a spatial 
resolution (i.e.~the grid point spacing for the innermost 
grid) of 0.016~AU.
As a final check on the numerical accuracy, we used 
images computed with the
ray-tracing procedure at a large number of different 
inclination angles
 to determine the total flux (integrated over all wavelengths)
as a function of  the inclination angle.
Integration of these fluxes over all inclination angles then gives
the total model luminosity, which finally can be compared to the input 
luminosity. For the geometrical models described below,
we find typical deviations of $\la 1\%$ in the case of spherical and
thick disk models,
increasing up to $\sim 8\%$ for the thinnest disk models 
(when the vertical thickness of the
inner disk gets similar to the numerical grid resolution).

Note that our radiative transfer code not only accounts for
extinction by and emission from dust grains, but also treats scattering by
dust grains; isotropic scattering is implicitly assumed. Although
the scattering efficiency of dust grains is quite low at mid-infrared
wavelengths, scattering of optical/UV photons in the region above and below
the disk plane can lead to significantly increased heating of the outer
disk regions; this can result in more extended thermal emission at
 mid-infrared wavelengths. If scattering is neglected (as is sometimes
done in other radiation transfer simulations), the extent of the emission
can be significantly underestimated.

Our radiative transfer modeling assumes the disk is passive
(i.e.~the temperature at each point is determined by the local radiation 
field) and does not take into account possible viscous disk heating by 
accretion. This is appropriate because the
stellar parameters and the estimated accretion rate (see \S 1)
suggest that accretion does not contribute more than $\sim 10\%$
to the total luminosity.

A self-consistent radiation transfer model should, of course, reproduce not 
only the observed visibilities, but also the broad-band spectral 
energy distribution (SED). We therefore constructed the SED of HR~5999
from the UV
to mm wavelengths from literature data (Hillenbrand et al.~1992;
Henning et al.~1994; Acke \& van den Ancker 2004). 
A complication arises from the presence of the companion 
Rossiter~3930, which has an angular separation  of only $1.4''$
from HR~5999. The fluxes for the SED points, which were
obtained from seeing-limited observations, therefore represent the
sum of the fluxes from HR~5999 and Rossiter~3930.
The adaptive optics study by Stecklum et al.~(1995) showed that 
the flux of Rossiter~3930 is considerably smaller than that of HR~5999
at optical and near-infrared wavelengths; they determined
$\Delta (V, J, H, K) \sim (4.6,\,3.0,\,3.1,\,3.6)$~mag.
At longer wavelengths, however, Rossiter~3930 may well contribute a 
larger fraction to the total flux, and it cannot even be excluded 
that it may dominate the total flux, as is the case for the several well-known
``infrared companions'' in binary T Tauri star systems
(e.g.~Chelli et al.~1988).
Although the latter possibility does not appear to be very likely
in the case of HR~5999
(since there is no strong increase in the flux contribution from
Rossiter~3930 from the optical to the near-infrared range),
all SED points longward of $2.2\,\mu$m can only be considered as upper limits
to the flux of HR~5999. 
In our radiation transfer modeling we therefore cannot require that the 
model SED matches the observed SED longward of $2.2\,\mu$m, but only that it 
does not exceed the observed SED.
We note that the companion does not affect our MIDI data of 
HR~5999 since the interferometric field-of-view is smaller
than the separation\footnote{To be sure, we inspected the acquisition images and found no
indication of another source in the MIDI field-of-view.}.

As input for the radiation transfer calculations one has to specify 
the luminosity and effective temperature of the
central radiation source (we used $L_{\rm bol} = 87\,L_\odot$
and $T_{\rm eff} = 7925$~K following van Boekel et al.~2005), 
the optical properties of the dust,
and the density structure of the circumstellar 
material. 
The dust composition around HR~5999 was recently investigated
by van Boekel et al.~(2005), who
found that the $8-13\,\mu$m spectrum is dominated by large
silicate (olivine) grains. 
Guided by their results, we used a dust model here that consists of
silicate and carbon grains with sizes of $1\,\mu$m.
The sublimation temperature of both grain species is taken to be 1500~K.
In the radiation transfer code, the destruction of grains through
sublimation is simulated by setting the number density of the
grains to zero, if the temperature of that particular grain species,
which is calculated simultaneously with the radiation field, exceeds the
sublimation temperature.

Model images were computed with the radiation transfer code
for those 15 wavelengths between $8\,\mu$m and $13\,\mu$m for which
the visibilities were determined from the MIDI data.
We then calculated visibility maps from the model images
and compared them to the measured visibilities for
HR~5999.
Since the orientation of the disk  
(i.e.~the angle between the disk axis and the north-south direction
on the sky, measured counter-clock wise and denoted here as the 
position angle $\phi$) is unknown, 
an $\chi^2$ fitting procedure was used to determine
the position angle that results in the best agreement between measured
and simulated visibilities. For this we used all 9 visibility curves,
each consisting of 15 wavelengths points, i.e.~135 individual data points.

\subsection{Models for the density distribution}

We consider here two different classes of models for the density distribution 
of the circumstellar material,  disk models, and spherical envelope
models.

\begin{figure*}
\parbox{18cm}{\hspace{2.5cm}\includegraphics[width=4.0cm]{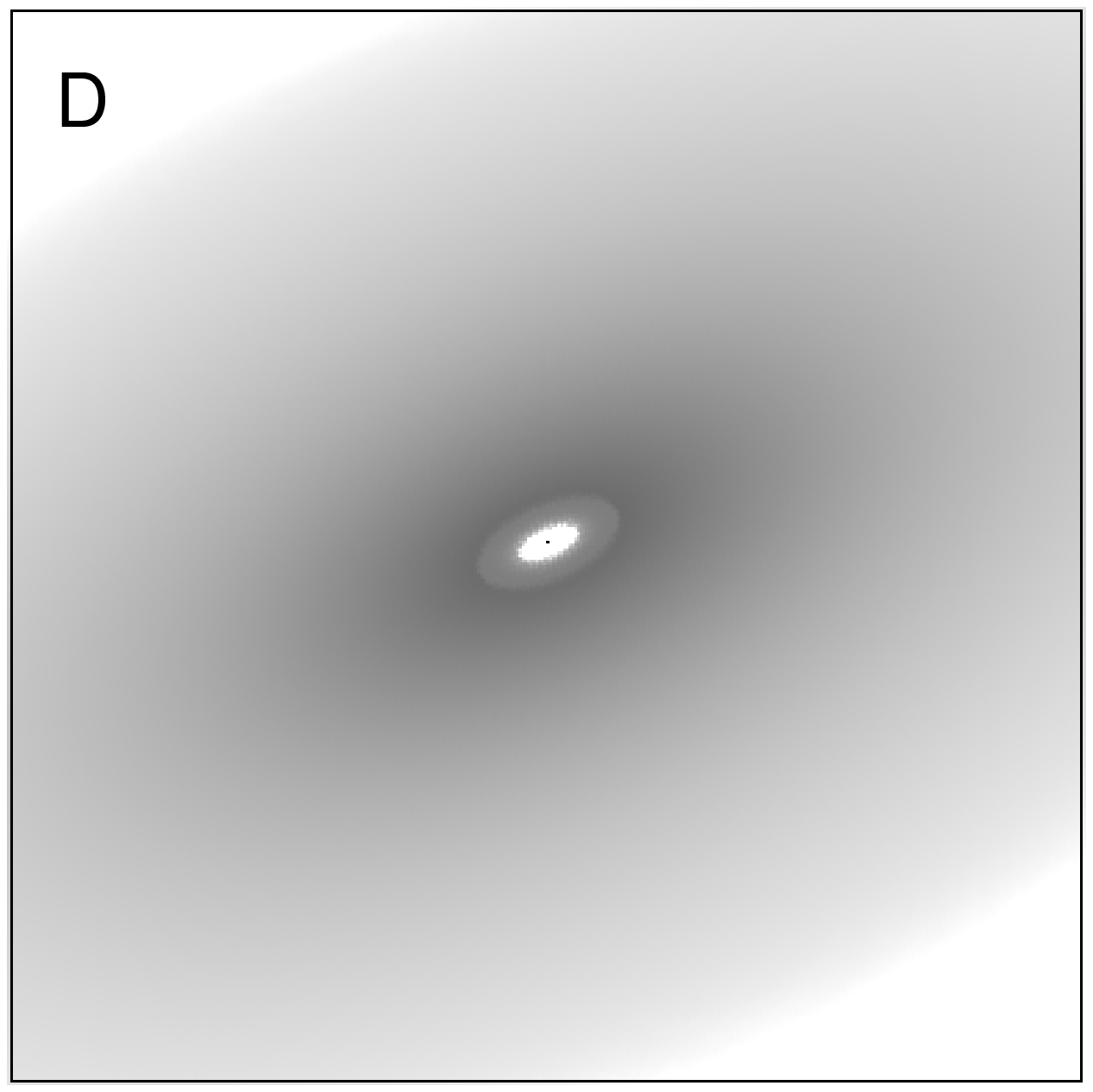}
\hspace{5mm}
\includegraphics[width=4.0cm]{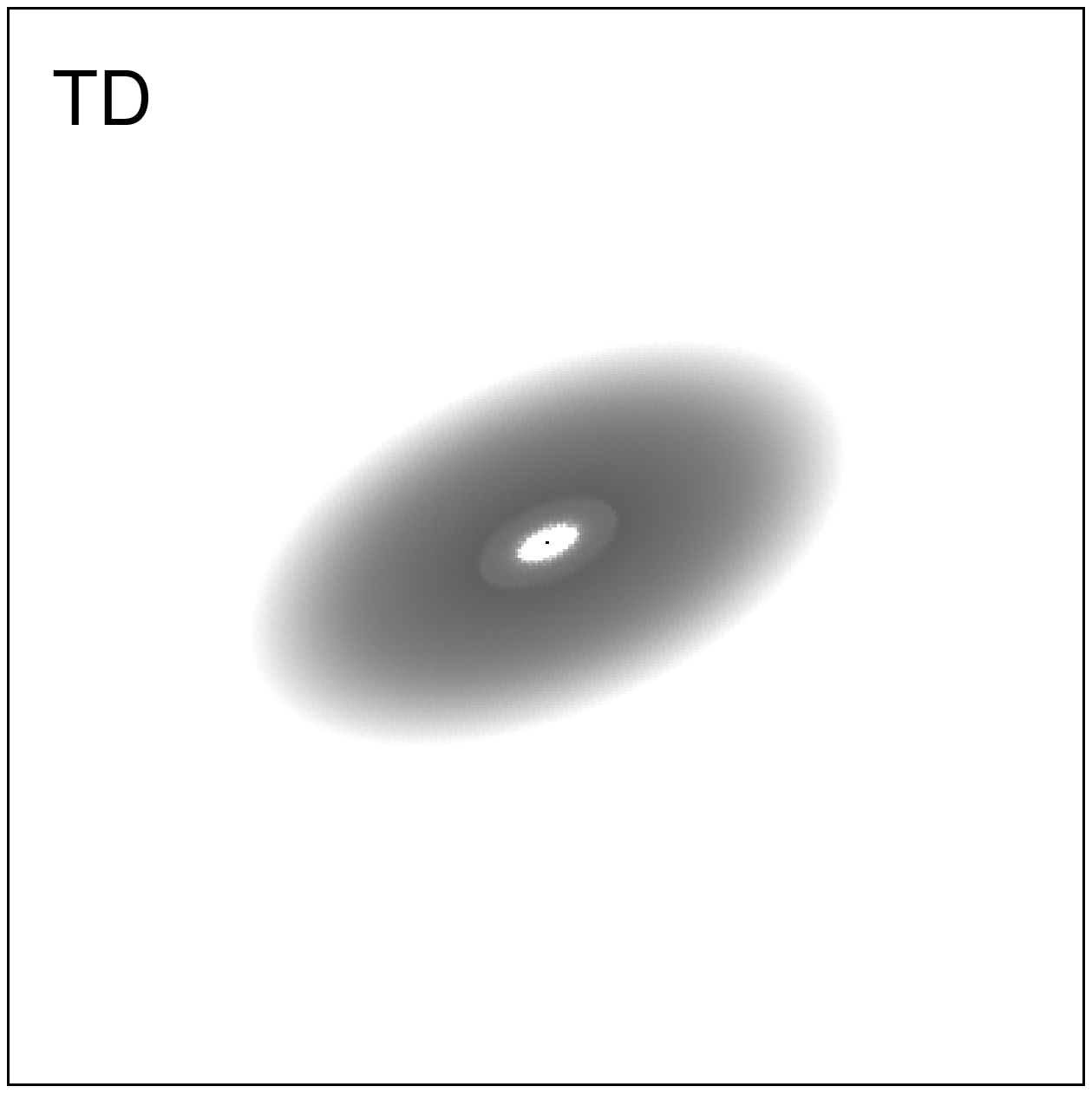}\hspace{5mm}
\includegraphics[width=4.0cm]{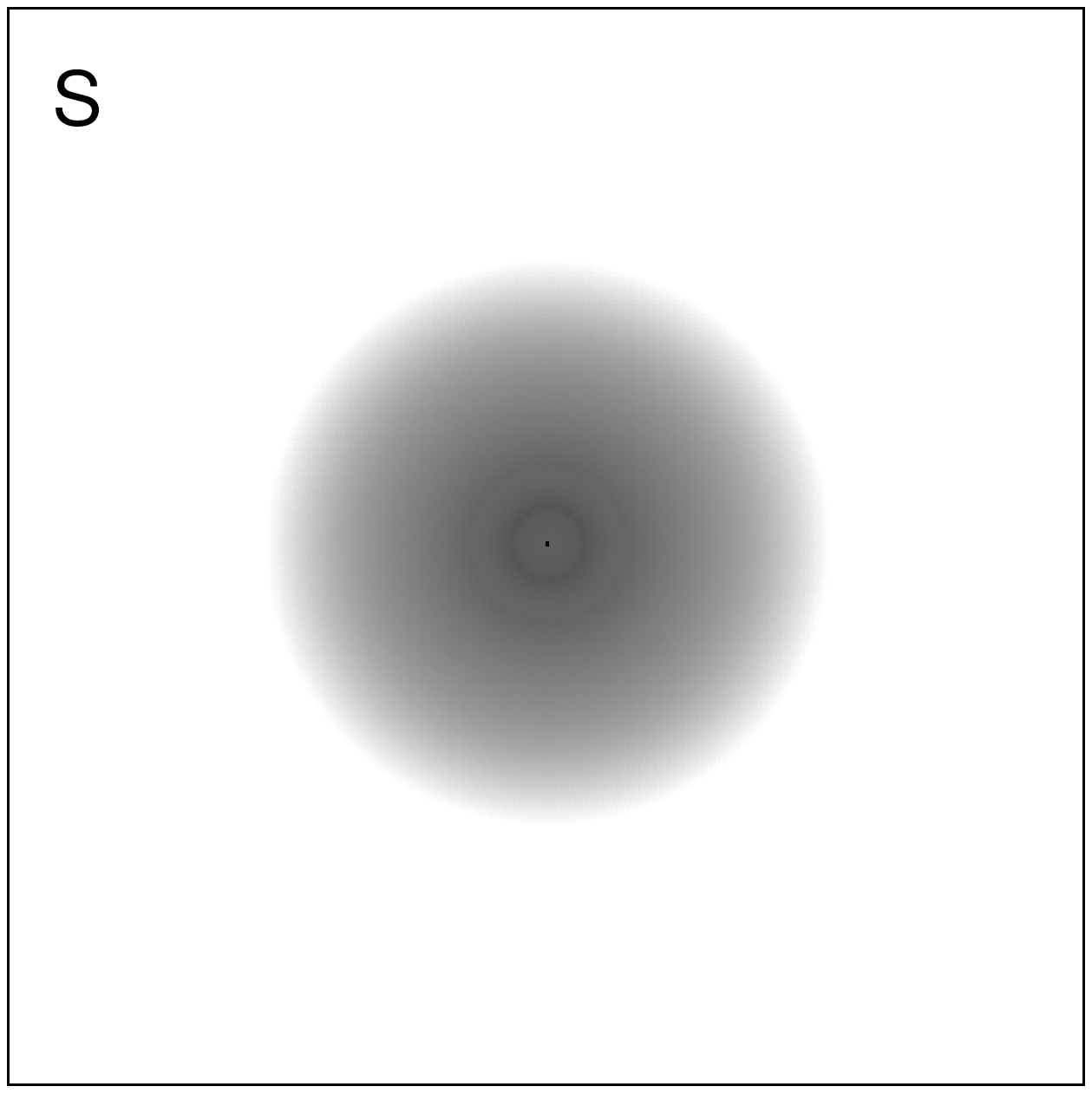}}
\includegraphics[angle=-90,width=6cm]{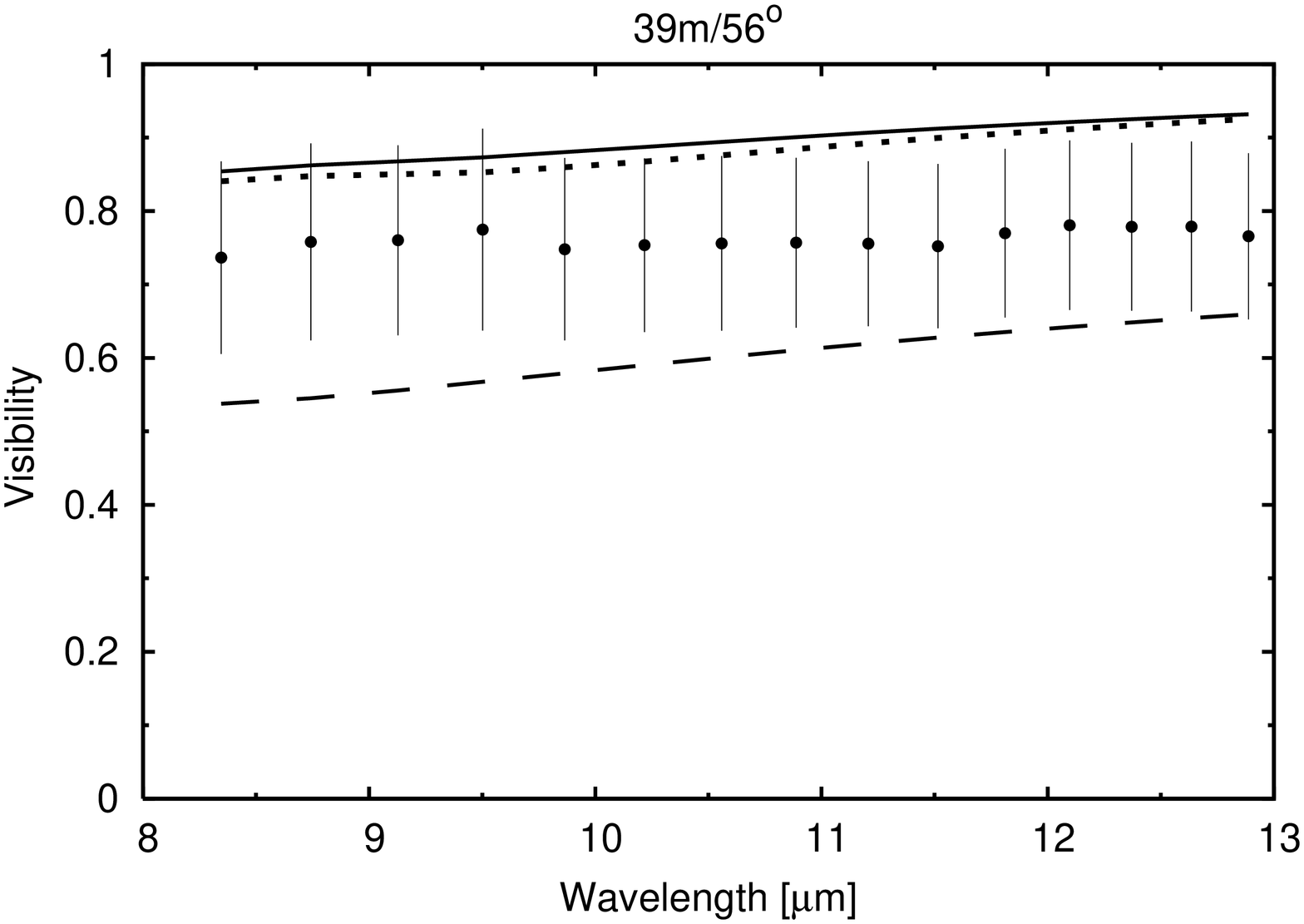}
\includegraphics[angle=-90,width=6cm]{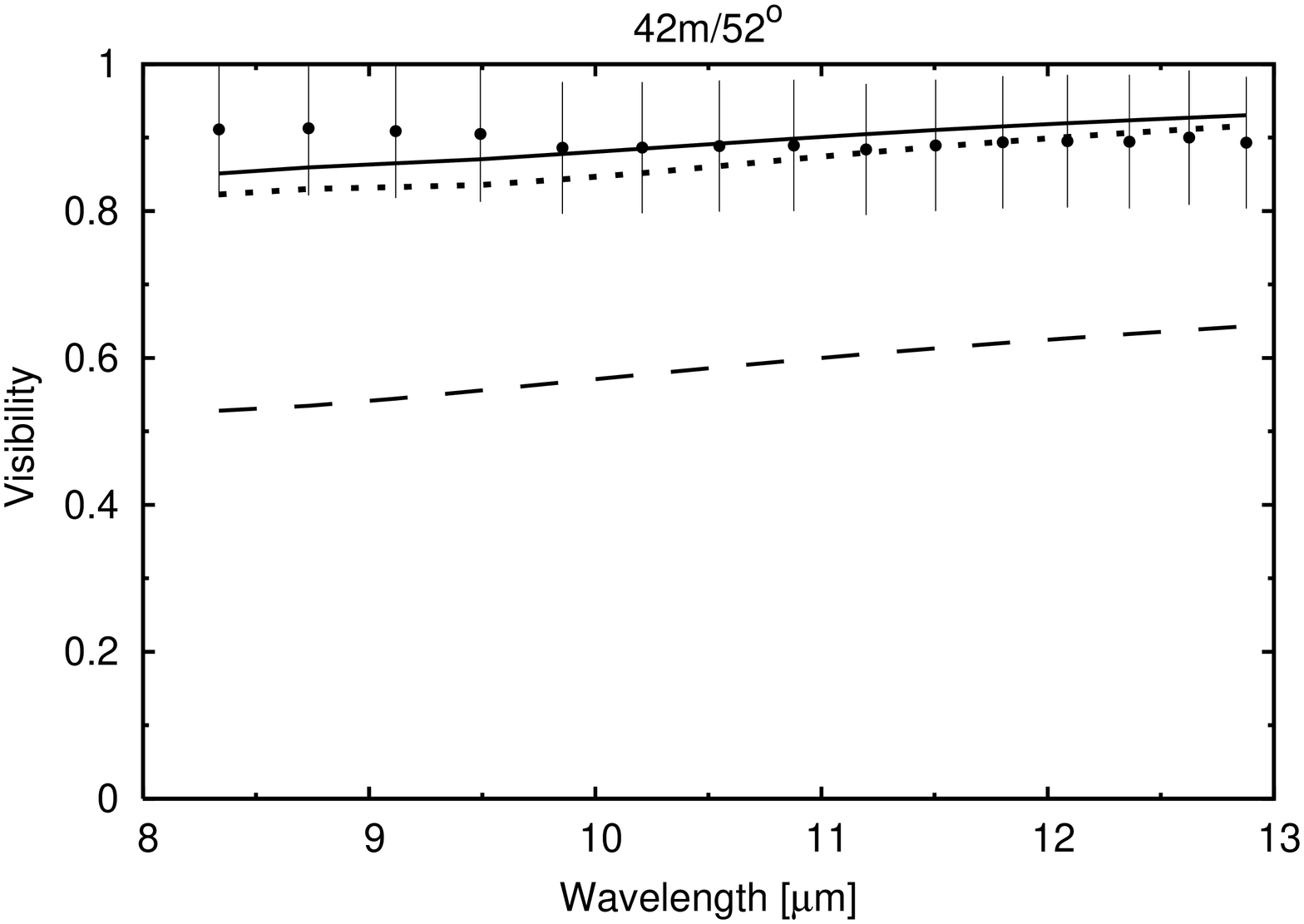}
\includegraphics[angle=-90,width=6cm]{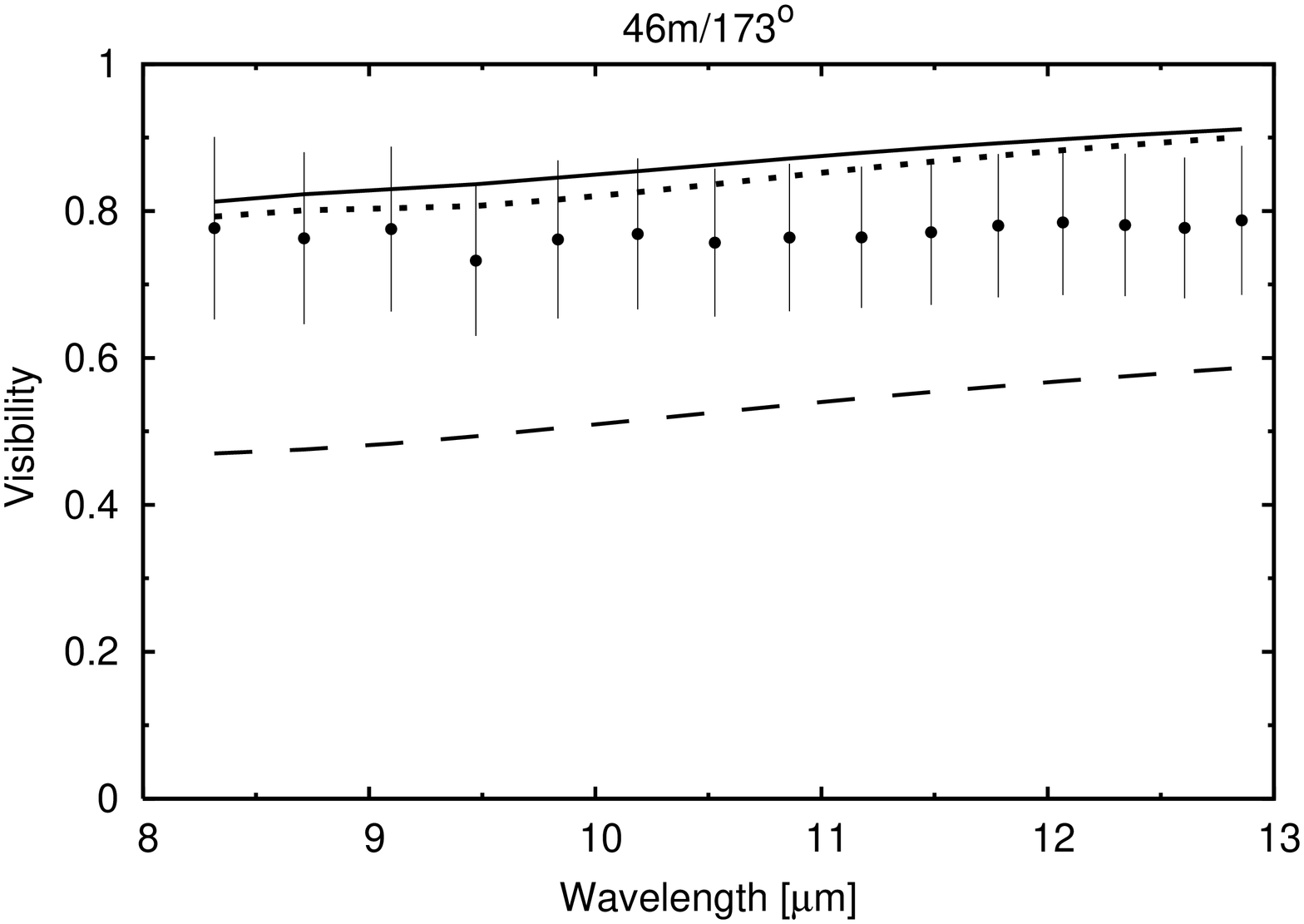}
\includegraphics[angle=-90,width=6cm]{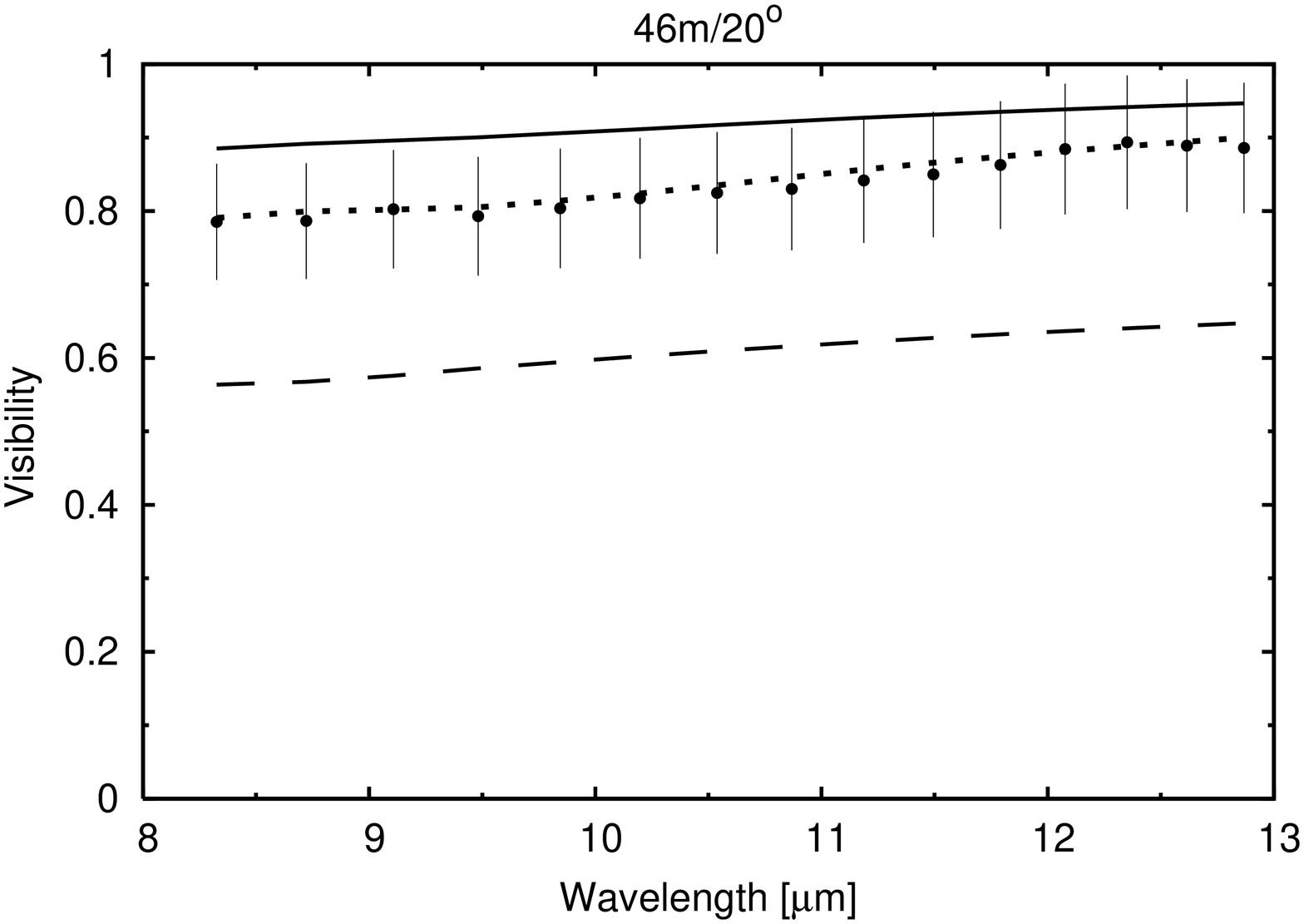}
\includegraphics[angle=-90,width=6cm]{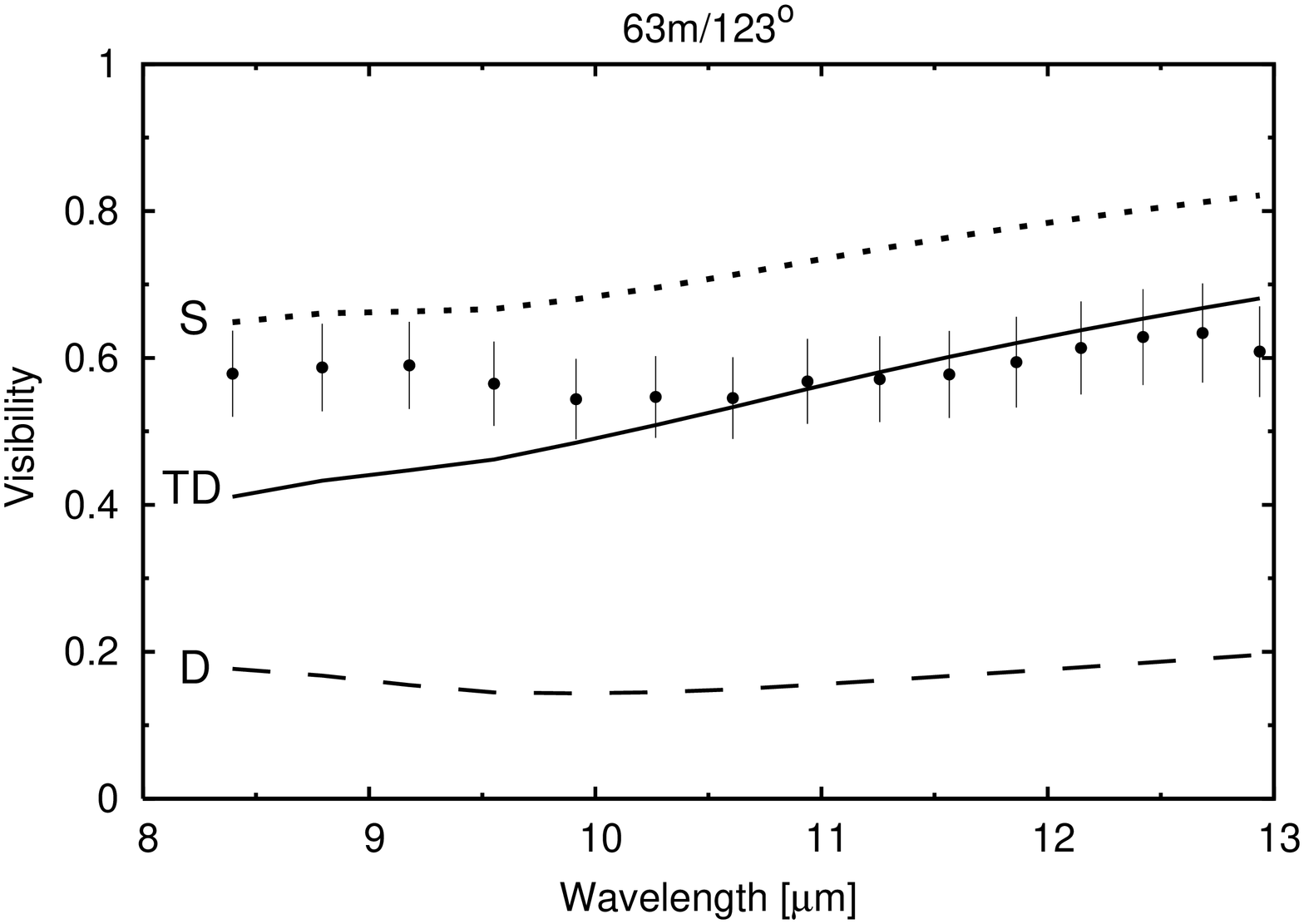}
\includegraphics[angle=-90,width=6cm]{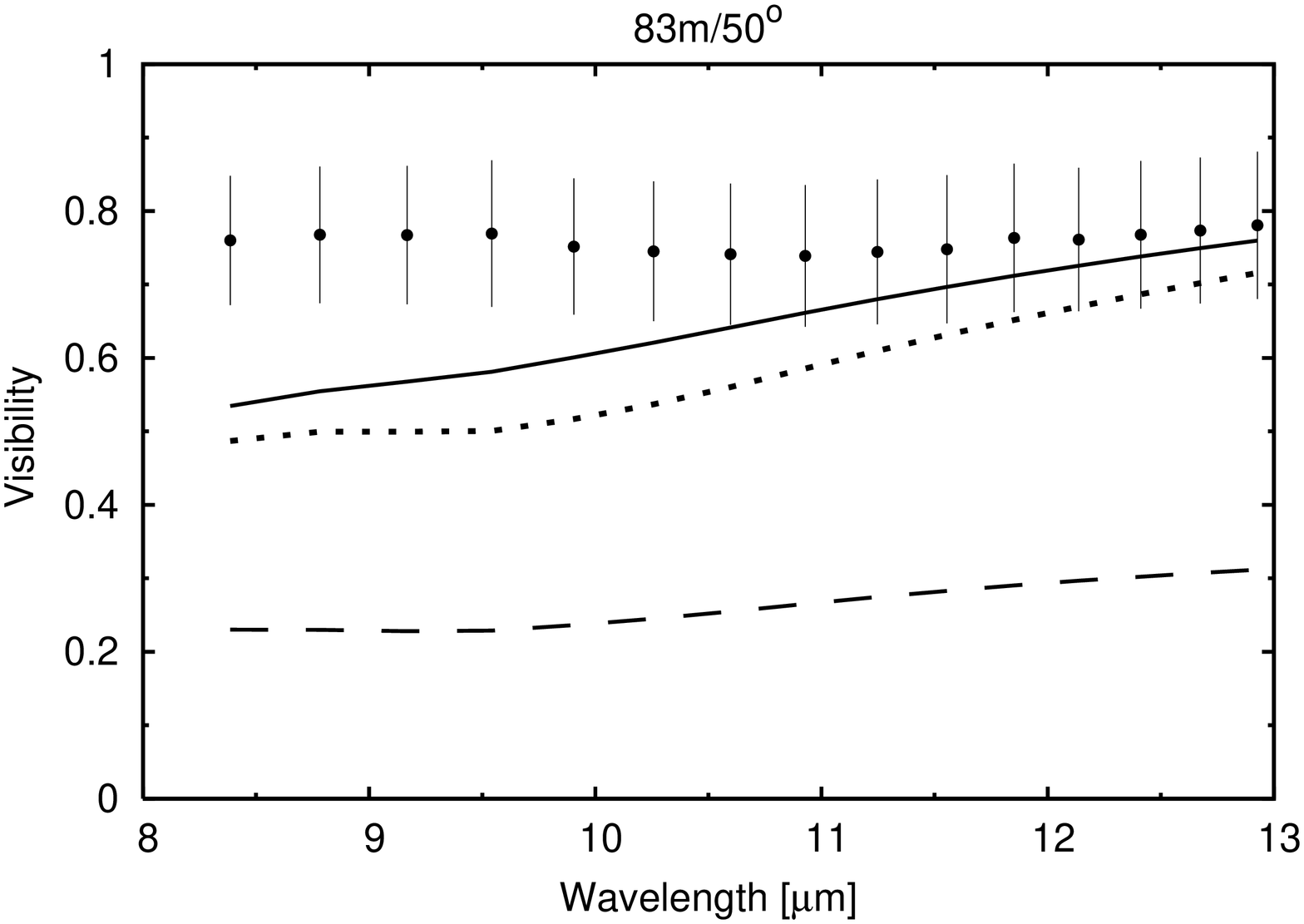}
\includegraphics[angle=-90,width=6cm]{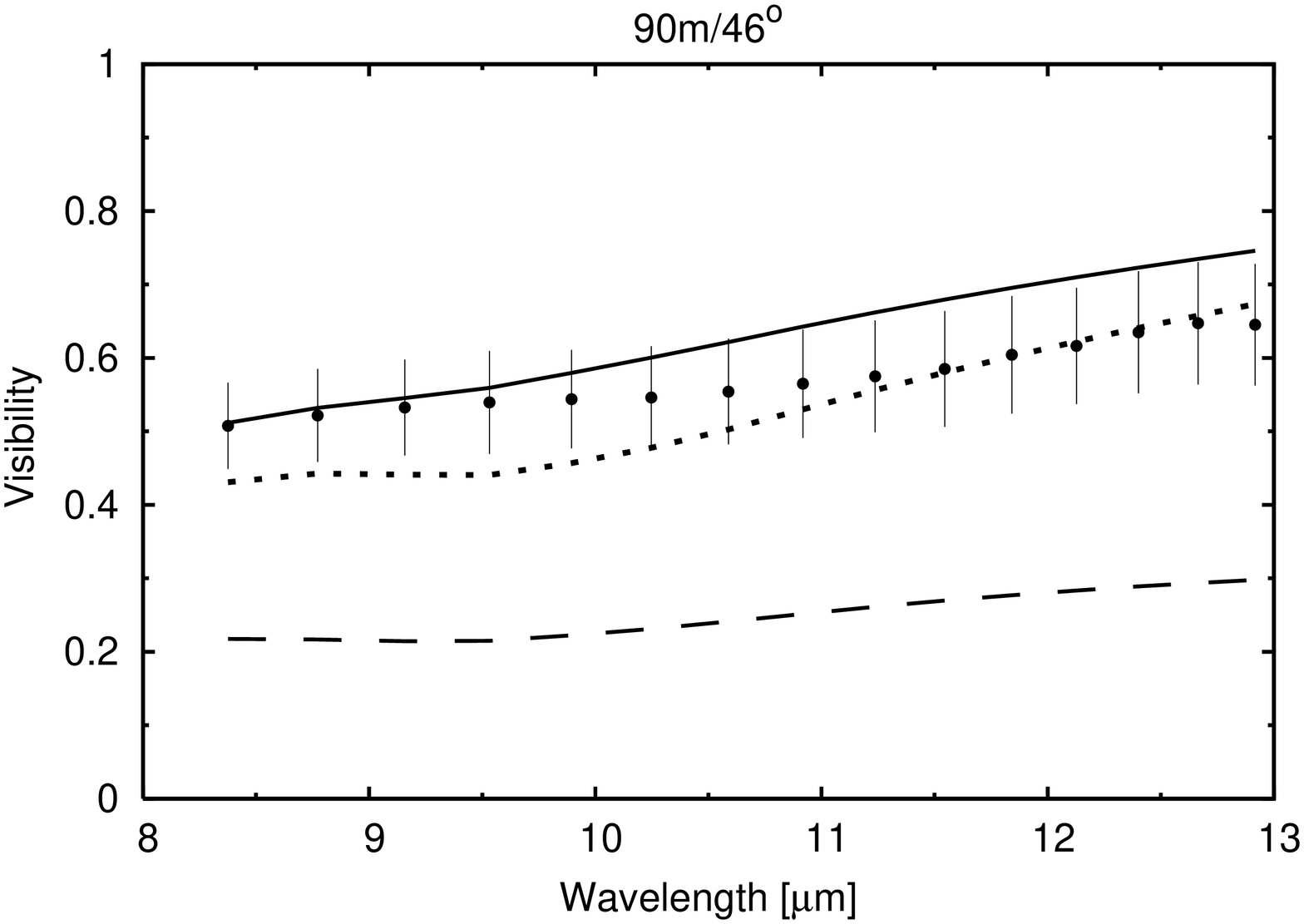}
\includegraphics[angle=-90,width=6cm]{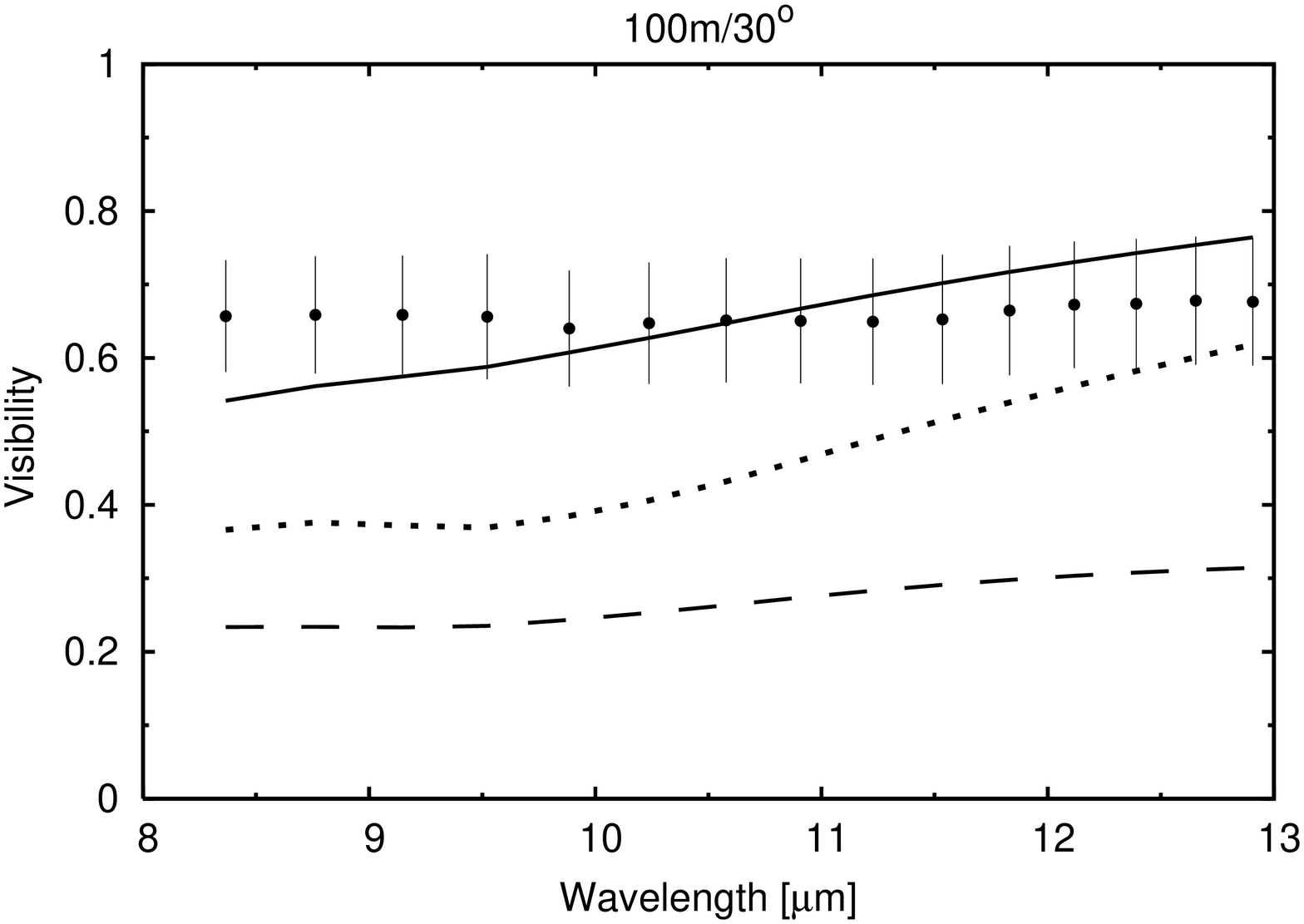}
\includegraphics[angle=-90,width=6cm]{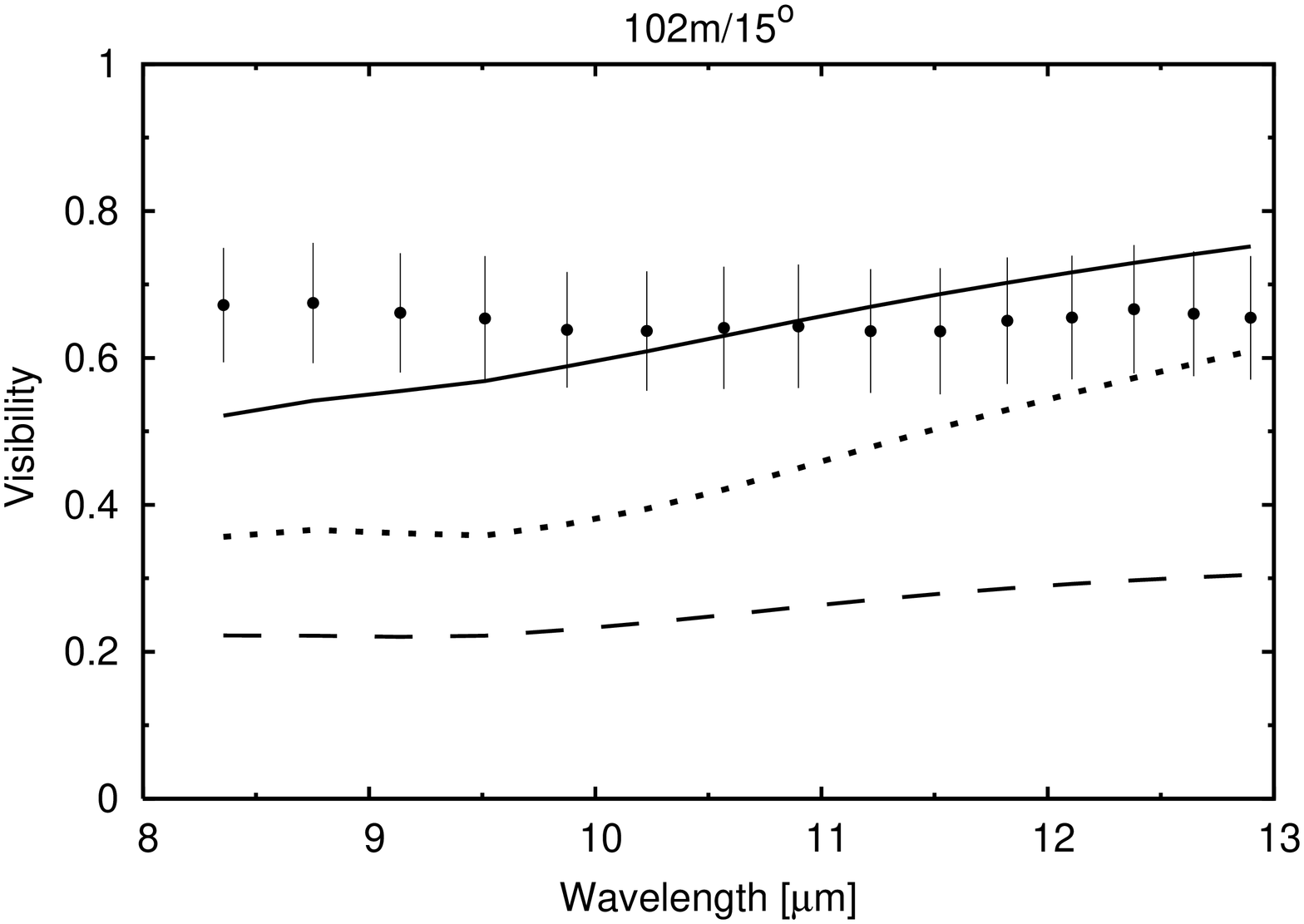}
\caption{
Radiation transfer modeling 
of HR~5999.
The upper row shows the model images at $10\,\mu$m for the disk
model ({\bf D}, left), the truncated disk model ({\bf TD}, middle), 
and the spherical
shell model ({\bf S}, right). Each image shows a $12.5\times 12.5$~AU 
region with a negative logarithmic intensity scale, seen under an
inclination angle of $i=59\degr$ and at  position angle $\phi = 24\degr$.
The  mosaic of visibility versus wavelength plots compares the 
observed visibilities of HR~5999 (solid dots with error bars)
for each individual MIDI observation
to the model visibilities; the truncated disk model ({\bf TD})
is displayed by the solid line, 
the disk model without truncation ({\bf D}) by the dashed line,
and the spherical shell model ({\bf S}) by the dotted line.
} \label{fits}%
\end{figure*}

\subsubsection{Extended disk (Model {\bf D})}

We started with models of
a quasi-Keplerian disk (i.e.~the disk structure is
determined by the gravitationally dominating central star),
with the following density distribution:
\begin{equation}
\rho_{\rm D}=\rho_0 \left(\frac{r}{r_0} \right )^{-p}\!\exp{\left[\D -\frac{\pi}{4}
\left (\frac{z}{h_z}\right )^2\,\right]},
\label{rho_kepler}\label{disk_mod}\end{equation}
$$
{\rm with}\;\;
h_z =  \left(\frac{r}{r_0} \right )^q \times h \cdot r_0. \hspace{3cm}
$$

We used $p = 15/8$
for the radial density power-law index, and for
the radial dependence of the vertical disk scale height $h_z$ we 
assumed $q = 9/8$, i.e.~a slightly flaring disk shape.
The parameter $h$ fixes the relative geometrical thickness of the disk.

In our modeling we 
considered different values for the relative disk thickness $h$ 
(from very thin disks with  $h = 0.05$ to very thick disks with 
$h \sim 0.8 $),
different  inclination angles $i$
(from $i=0\degr$ for a face-on view and $i = 90\degr$ for an edge-on view),
and also tried different values for the radial density power-law index $p$.
For each combination of model parameters, the density $\rho_0$ 
is restricted by the requirement that the model SED should agree with the 
observed SED points for wavelengths $\lambda \leq 2.2\,\mu$m
and must not exceed the observed fluxes for longer wavelengths
(where the observed SED points contain an unknown flux contribution from
the binary companion).

The resulting brightness distribution in the simulated images 
is dominated by the central star and the inner disk region of a few AU 
(see Fig.~\ref{fits}). The peak brightness is found at the
inner rim of the dusty disk at the dust sublimation radius, and it drops 
smoothly outwards.
Despite the quite compact appearance of the simulated images, 
the weak emission from the outer parts of the disk
contributes a significant fraction of the total flux at mid-infrared
wavelengths.
This extended emission leads to a large effective size of the emission 
and yields
model visibilities that are considerably lower 
than some of the observed visibilities.
The characteristic size of the model brightness distributions is up 
to twice as large as 
the sizes derived from the observed visibilities.
Even if we change the power-law exponent for the radial density law
(to values as steep as $p = 4$) or the thickness of the disk, 
the extent of the emission in the 
model images is still considerably
too large and inconsistent with the observed visibilities.

\subsubsection{Truncated disk (Model {\bf TD})}

One possibility for decreasing the amount of
extended emission in the outer parts of the images is 
to limit the circumstellar mass distribution to a certain outer radius. 
For this, we used the disk model describe above (Equ.~\ref{disk_mod})
and truncated the density distribution at the outer radius $r_t$.
This truncation is performed by multiplying
the density distribution with a Fermi-type function:
\begin{equation}
\rho_{\rm TD} =\rho_{\rm D} \times
\left ( 1+\exp\left[\frac{R-R_t}{\alpha}\right]\right)^{-1}.
\label{rho_trunc_kep}\end{equation}
where $\rho_{\rm D}$ is the density distribution given in Equ.~(\ref{rho_kepler}), 
$R = \sqrt{r^2 + z^2}$ is the radial distance from
the origin, while
$\alpha$ determines the width of the transition zone of the
Fermi-type function; we used $\alpha = R_t/10$.

\begin{figure}
\begin{center}\includegraphics[width=7.5cm]{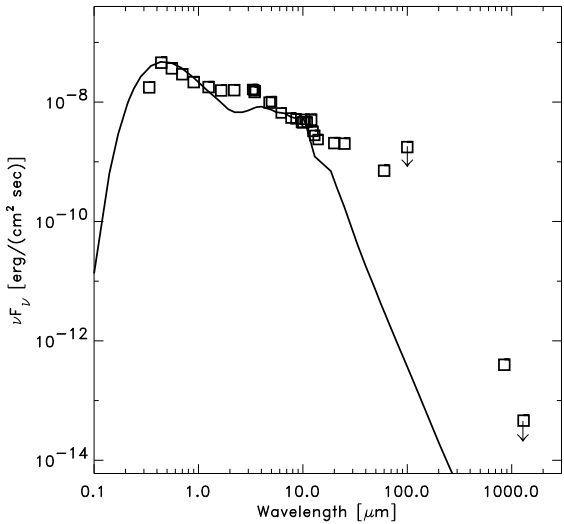}\end{center}
\caption{Observed spectral energy distribution (SED)
of HR~599 (boxes; compiled from the references listed in the text)
compared the predicted SED for the 
truncated disk model described in the text.
While the model reproduces the SED reasonably well up to
$\lambda \sim 10\,\mu$m, it
does not account for the bulk of the cool dust responsible for
the emission at $\ge 13\,\mu$m. This cool dust may be associated to the
companion star Rossiter~3930 or perhaps 
widely distributed in the HR~5999/R~3930
system. (Note that dust at distances $\ge 100$~AU from HR~5999
would be cool enough to produce excess emission only longward of $10\,\mu$m.
The beamsize of the far-infrared SED points corresponds to sizes of several
1000 AU, and the data do therefore not allow to constrain the location of the
cool dust.)
} \label{sed}%
\end{figure}

With these models we were able to find
acceptable fits to the observed SED up to $\lambda \sim 10\,\mu$m 
(see Fig.~\ref{sed}) and the observed visibilities
for truncation radii in the range $\sim  2 - 3$~AU,
inclination angles of $i \sim 60\degr$, and position angles of $\phi \sim 25\degr$.
The vertical thickness of the disk (governed by the parameter $h$)
is nearly unconstrained by the data (thin and moderately thick
disks seen under intermediate
inclination angles ``look'' very similar).

In Fig.~\ref{fits} we show a model with parameters $h = 0.056$,
$R_t = 2.6$~AU, $i = 59\degr$, and $\phi = 24\degr$, which yields 
an acceptable (albeit not perfect) fit to the data with
$\chi^2_r = 1.07$. The V-band optical depth of this model along the
disk plane is $\tau = 3.5$.
Although models with larger vertical disk scale height 
give marginally better fits to the visibilities (for example, $\chi^2_r = 0.98$
for a disk with $h = 0.4$), we prefer this specific thin-disk model because the 
vertical scale height of the dusty disk
agrees with the formal condition of vertical
hydrostatic equilibrium (see, e.g., Equ.~7.
in Chiang \& Goldreich 1997).
 
For comparison, we note that
a model for the same disk parameters but without the truncation factor
(model {\bf D} in Fig.~\ref{fits})
yields $\chi^2_r = 18.3$.

\subsubsection{Spherical shell (Model {\bf S})}

As already shown in \S \ref{shape}, the observed visibilities suggest
that the brightness distribution is not spherically symmetric but rather
somewhat elongated.
In order to assess quantitatively how much the truncated disk model 
provides a better description to the data than spherical shell models,
we also considered models with a spherically symmetric
power-law density distribution, in which the density distributions were
again truncated at radius $R_t$:

\begin{equation}
\rho_{\rm S}=\rho_0 \left(\frac{r}{r_0} \right )^{-p} \times
\left ( 1+\exp\left[\frac{R-R_t}{\alpha}\right]\right)^{-1}
\label{rho_shell}\end{equation}

We found that spherical shell
models always provide poorer fits to the data than the truncated disk models.
A spherical model (model {\bf S} in Fig.~\ref{fits})
with the same radial density power-law 
and truncation radius as in the truncated disk model
described above yields $\chi^2_r = 2.91$. 

\subsection{Conclusions from the radiative transfer modeling}

We find that dust distributions in which the radial density follows a
power law (with exponents in the range $p \sim 1 - 4$) 
produce brightness distributions that are much too
extended to be consistent with our set of MIDI data.
Disk models in which the density is truncated at outer radii of 
$\sim  2 - 3$~AU, on the other hand, provide reasonably 
good fits to the MIDI data. 

Given the large number of free parameters in the radiative transfer models
and the relatively large visibility errors, which do not allow us to 
constrain model parameters very well, it was not our aim to scan
the full multi-dimensional parameter space in order 
to find the overall ``best fit''
model. We instead proceed the other way round and summarize here those model
categories that can be ruled out by our data and modeling.
The following cases are inconsistent with the MIDI data:
\begin{itemize}
\item {\bf Spherical shells --\,} The 
 observed visibilities suggest that the brightness distribution is not 
 circular symmetric but rather somewhat elongated. This is confirmed
by our radiative transfer
modeling, which shows that spherically symmetric shell models cannot produce
good fits to the observed data.
\item {\bf Disks seen at low inclination angles ($i < 30\degr$, i.e.~nearly
``face-on'') --\,} Such models produce a nearly circular symmetric 
brightness distribution and thus do not fit the data well.
\item {\bf Very thick disks ($h > 0.7$) --\,} Such models also
produce only very slightly elongated, nearly circular symmetric brightness 
distributions.
\item {\bf Geometrically thin disks ($h < 0.2$) 
seen close to edge-on ($i \ge 75\degr$) --\,}  Such models predict strongly
elongated brightness distributions, which are inconsistent with the moderate
amount of asymmetry suggested by the MIDI data.
\end{itemize}

Finally, we
note that there may be other ways to explain the compact
brightness distribution. One possibility for decreasing the amount of
extended emission in the outer parts of the images may be
to introduce a dense puffed-up wall at the inner edge of the
dust distribution that casts a shadow on the outer parts of the disk.
Such a shadow leads to a much steeper temperature profile in the disk
regions behind the rim, and therefore the region containing warm enough 
dust that radiates significantly in the mid-infrared is smaller than 
without  an inner rim.
Models of this kind have, for example, been used by Dullemond et al.~(2001)
 to fit the observed SEDs of HAEBEs. The resulting brightness distribution
will depend on the detailed shape of the inner rim,
which is not well known (see, e.g., discussion in Isella \& Natta 2005).
Since such models introduce several additional free parameters for the 
radiative transfer modeling, 
they are beyond the scope of the present study and will be
the topic of a forthcoming paper.

\section{Summary and conclusions}

Our MIDI interferometric study has, for the first time, resolved
the mid-infrared emission around the Herbig Ae star HR~5999.
The most remarkable result 
is the compactness of the $10\,\mu$m emission, which has a characteristic
radius of only $\sim 2-3$~AU.
The data also suggest that the emitting region is elongated.
Our radiative transfer modeling shows that  
a moderately thick disk with outer edge at $\sim 2.6$~AU,
seen under an intermediate inclination angle, provides a
reasonably good description of the data. We also find that
disk models provide better fits to the data than
spherical shell models.

The derived outer disk radius of $\sim 2.6$~AU is considerably smaller than 
canonical assumptions of disk sizes around young stellar objects.
This leaves us with the question of why the disk around HR~5999 is so
compact.
One possible reason may be that the disk was truncated by the 
gravitational effect of a companion to HR~5999. Dynamical disk
clearing is expected to truncate the disks around the stars in a  binary
system to roughly $1/3$ of the orbital separation of the components
(see, e.g., the numerical simulations by Pichardo et al.~2005).
The presence of the companion Rossiter 3930 is interesting in this context,
but the quite large projected separation
between HR~5999 and Rossiter~3930 of $\sim 300$~AU would require an
extremely elliptical orbit in order to explain a disk truncation radius 
of only $\sim 3$~AU.
A more likely possibility would be the presence of another, yet undetected, 
companion at close separation ($\sim 10$~AU) from HR~5999.
The MIDI data provide no hint of the existence of a close companion,
but we note that it would essentially be undetectable in the MIDI data
if its mid-infrared emission\footnote{ We performed a number of simulations
to get a realistic estimate of the MIDI 
detectability of a possible companion in the following way:
To our "best fit" model images (for the truncated disk model)
we artificially added a "companion" at a separation of 50 mas from the
central star. 
We considered a range from  1:100 to 1:1 for
the flux ratio between the emission from the
companion and the photospheric emission from the primary star.
We then computed the wavelength-dependent visibilities 
for the  $(u,v)$ points of our MIDI observations from these images, and
finally determined the difference to the visibilities without the simulated
companion. As a general result, we found that the
effect of the simulated companion on the visibilities is very small.
Even in the case of a very bright companion (flux ratio 1:1), 
the maximal change in the visibilities was less than 3\%,
i.e. well below the uncertainty of the MIDI visibilities.
We note that, for example, a $1\,M_\odot$ companion with an age of about
1~Myr would yield a photospheric flux ratio of 1:9.
Thus, even if the companion  had mid-infrared excess due to its own disk
at a level of 10 times the photospheric flux, it would be undetectable
in the MIDI data.
}
 is less than a few percent of that from
the material around HR~5999.

\begin{acknowledgements}
We would like to thank
the referee for detailed comments that helped
to improve this paper. 
\end{acknowledgements}

\end{document}